\documentclass[twocolumn,superscriptaddress,longbibliography,aps,pra,preprintnumbers]{revtex4-2}

\usepackage[urlcolor=blue,colorlinks=true,citecolor=blue,linkcolor=blue,pdfstartview={FitH},bookmarks=false]{hyperref}
\usepackage[utf8]{inputenc}
\usepackage{graphicx,comment}
\usepackage{xcolor,soul}
\usepackage{dcolumn}
\usepackage{bm}
\usepackage{amsmath}
\usepackage{amssymb}
\usepackage[normalem]{ulem}

\graphicspath{{fig/}{./fig/}{.}}

\begin{document} 

\title{Insulating Half-Heusler TmPdSb with Unusual Band Order\\ and Metallic Surface States}

\author{Shovan Dan}
\email[e-mail: ]{dan.shovan@gmail.com}
\affiliation{Institute of Low Temperature and Structure Research, Polish Academy of Sciences, Okólna 2, 50-422 Wrocław, Poland}

\author{Andrzej Ptok}
\email[e-mail: ]{aptok@mmj.pl}
\affiliation{Institute of Nuclear Physics, Polish Academy of Sciences, W. E. Radzikowskiego 152, PL-31342 Kraków, Poland}

\author{O. Pavlosiuk}
\affiliation{Institute of Low Temperature and Structure Research, Polish Academy of Sciences, Okólna 2, 50-422 Wrocław, Poland}

\author{Karan Singh}
\affiliation{Institute of Low Temperature and Structure Research, Polish Academy of Sciences, Okólna 2, 50-422 Wrocław, Poland}

\author{P. Wiśniewski}
\affiliation{Institute of Low Temperature and Structure Research, Polish Academy of Sciences, Okólna 2, 50-422 Wrocław, Poland}

\author{D. Kaczorowski}
\affiliation{Institute of Low Temperature and Structure Research, Polish Academy of Sciences, Okólna 2, 50-422 Wrocław, Poland}

\date{\today}

\begin{abstract}
We present theoretical and experimental results exploring a half-Heusler compound TmPdSb with unusual band order and metallic surface states.
Typically, the half-Heusler systems exhibit topological features in a semimetallic state, and trivial ones in an insulating state. 
Topological properties of the most of half-Heusler systems are related to the band inversion around the Fermi level, similar to this observed in CdTe/HgTe systems.
In the case of TmPdSb we observed the gapped electronic band structure with ``band inversion'' in the conductance band, while the slab-like system realized metallic surface states.
The bulk insulating nature of the compound was corroborated by means of electrical transport measurements. 
The experimental data revealed several features due to the presence of metallic surface states, such as linear magnetoresistance and weak-antilocalization effect, characterized by enhanced coherence length and a very large number of surface conductive channels.
Our findings reveal new features of the rare-earth half-Heusler. 
\end{abstract}

\maketitle


\section{Introduction}

The half-Heusler (HH) compounds are a wide class of materials possess a lot of interesting physical properties that make them attractive for various applications in thermoelectrics, spintronics, and other fields.
For example, the HH exhibits excellent thermoelectric properties~\cite{quinn.bos.21,zhu.li.23,dong.luo.22}, making them promising candidates for waste heat recovery and solid-state cooling applications.
Electronic properties can be relatively simply tuned by e.g. doping~\cite{dong.luo.22,chadov.qi.10,nakajima.hu.15}, which allows tailoring their electrical conductivity, topological properties, and other electronic characteristics for specific applications.
Certain HH compounds possess unique magnetic properties, making them suitable for spintronics applications~\cite{ma.hegde.17}.
HH compounds typically exhibit high thermal stability, enabling them to withstand high temperatures without significant degradation~\cite{xing.liu.19}, which opens the way to their applications in hard environments or high-temperature conditions.
Also, mechanical properties are suitable for structural applications due to their high hardness and strength~\cite{chen.ren.13,rogl.grytsiv.16}.
The unique combination of electronic, magnetic, thermal, and mechanical properties in HH compounds makes them promising materials for a wide range of technological applications.

In HH compounds, one can observe insulating~\cite{yan.devisser.14}, semiconducting~\cite{shi.si.17}, or semimetallic~\cite{armitage.male.18} features.
A theoretical study suggests that some of them can have non-trivial topology of electronic structure~\cite{lin.chen.15}.
Indeed, a precise investigation of the band structure~\cite{zhu.cheng.12,feng.xiao.10,chadov.qi.10,lin.wray.10} predicted a band inversion in some HH compounds~\cite{chadov.qi.10,lin.wray.10,feng.xiao.10, alsawai.lin.10,nakajima.hu.15,souza.crivillero.23}, and in consequence the unusual surface states (SS)~\cite{souza.crivillero.23,liu.lee.11,liu.yang.16,hosen.dhakal.20}. 
Topological properties are manifested in experiments as anomalous Hall effect~\cite{suzuki.chinsell.16,shekhar.kumar.18,singha.roy.19,orest.20,zhu.singh.20,zhang.zhu.20,chen.li.21,chen.xu.21,chen.li.22}, large negative magnetoresistance induced by chiral anomaly~\cite{YbPtBi,hirschberger.kushawaha.16,chen.li.20}, or negative shift in $^{209}$Bi nuclear magnetic resonance~\cite{nowak.pavlosiuk.15,zhang.hou.16}.

Typically, the electronic band structure of HH compounds is very similar to those of CdTe/HgTe~\cite{bernevig.hughes.06}.
In the case of topologically trivial CdTe (and insulating HHs with direct gap), the band structure at $\Gamma$ point realized ``normal'' order $\Gamma_{6}$--$\Gamma_{8}$--$\Gamma_{7}$ (from higher to lower energies).
Contrary to this, in topological non-trivial HgTe (and semimetallic HHs), the band order is changed drastically to $\Gamma_{8}$--$\Gamma_{6}$--$\Gamma_{7}$. 
The $p$-type quartet in $\Gamma_{8}$ state is protected by the cubic symmetry.
Whereas, the position of the $s$-type doublet in $\Gamma_{6}$ plays a crucial role in the band inversion and the topological properties of HH compounds.
However, in the case of the HH system with heavier rare earth atoms, this scenario cannot be realized~\cite{yan.devisser.14}.

Typically, the SS can be acclaimed by angle-resolved photoemission spectroscopy (ARPES) or scanning tunneling microscopy (STM), which both require cleavable single crystals. 
However, indirectly, the presence of SS has been argued from the magneto-transport properties of the material~\cite{Binghai}. 
Linear in magnetic field ($B$) magnetoresistance (MR) and/or quantum oscillations were proclaimed to  originate from the SS~\cite{Zhang.12, Qu.10}. 
In contrast to the SS, the presence of Weyl nodes in the bulk band structure requires the breaking of inversion symmetry and/or time-reversal symmetry~\cite{armitage.male.18}. 
Crystal structure of HH compounds lacks inversion symmetry and the presence of a magnetic order can break time-reversal symmetry. 
A few members of the HH class of material, viz, GdPtBi, NdPtBi, LuPdBi, ScPtBi, DyPdBi, YbPtBi etc. are reported from direct or indirect evidence to have Weyl nodes~\cite{shekhar.kumar.18, LuPdBi,ScPtBi,DyPdBi,YbPtBi}. 
Nevertheless, such exotic behavior originates mainly from the strong spin-orbit-coupling (SOC). 
The strong SOC also manifests in the conductivity with weak localization (WL) or weak antilocalization (WAL) due to the quantum interference. Interestingly, the coherence length also provides an idea of the topological nature of the compound~\cite{Bi2Te3,Minhao}.

\begin{figure*}
\centering
\includegraphics[width=\textwidth]{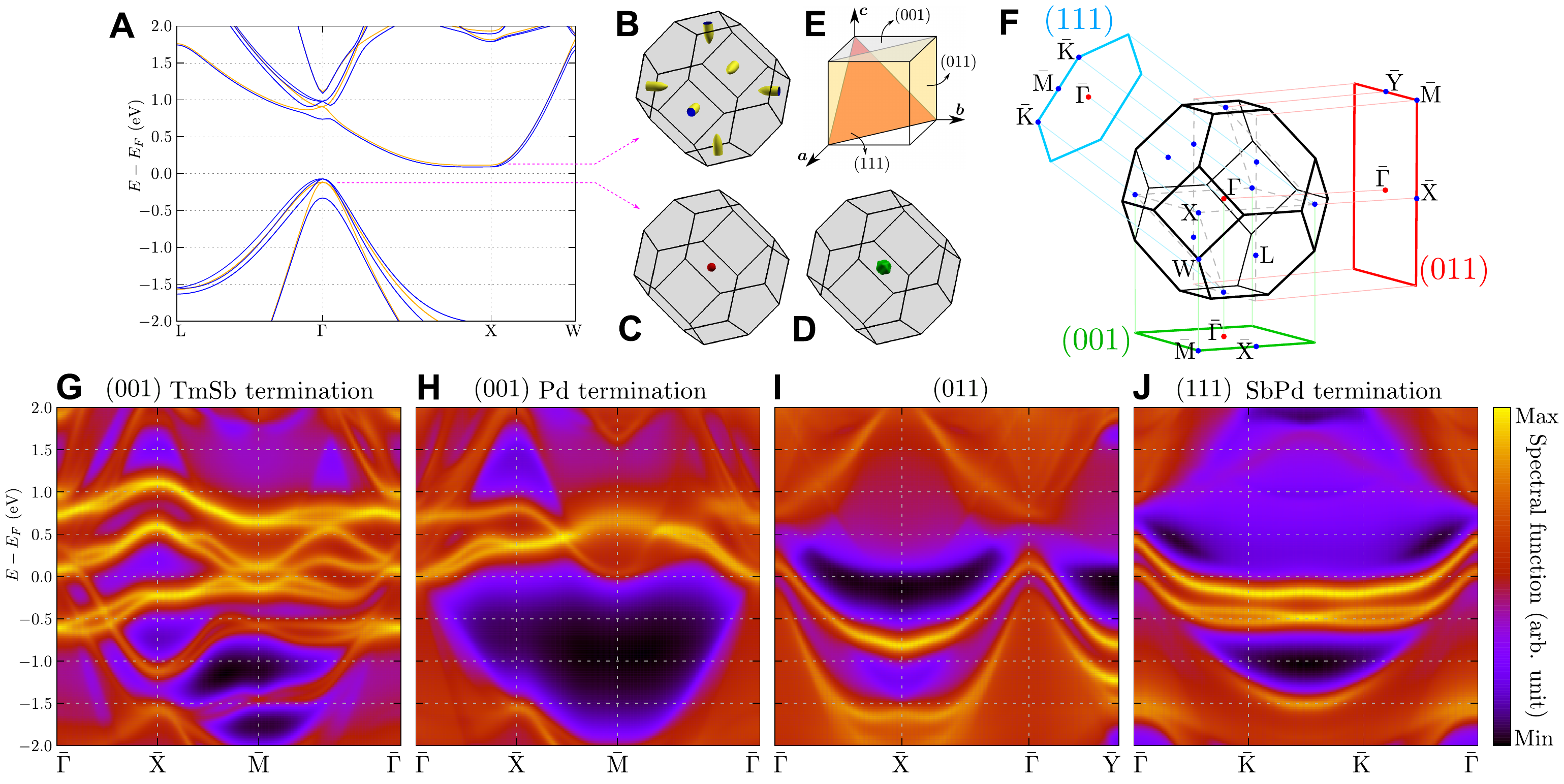}
\caption{
{\bf Results of theoretical calculations.}
({\bf A}) The electronic bulk band structure, resulting from the meta-GGA mBJ DFT calculation. The orange and blue lines represent bands without and with spin--orbit coupling, respectively.
({\bf B-D}) The Fermi surfaces of TmPdSb, with the Fermi level shifted by $\sim 0.1$~eV with respect to the bulk band gap (level marked by the pink dashed lines on ({\bf A})).
Panel ({\bf B}) correspond to the unoccupied band, while ({\bf C,D}) to occupied bands.
({\bf E}) Possible surface orientations of cubic structure.
({\bf F}) The three-dimensional bulk Brillouin zone and its projection on the surface Brillouin zones (for different orientations presented on ({\bf E})), high symmetry points are also engraved.
({\bf G-J}) Theoretically investigated surface spectral function for the different surface orientations and terminations (as labeled).
Surface (011) is terminated by all atoms contained in TmPdSb.
In the case of (111) with TmPd termination (not presented), the surface states are not realized.
\label{fig:band}
}
\end{figure*}

Here, we discuss the magneto-transport properties of TmPdSb, as a HH topological insulator candidate.
Our experimental results are supported by the {\it ab initio} calculations, which directly show that TmPdSb is an insulator with an indirect gap, characterized by unusual band inversion, and metallic surface states.
Such properties can draw new light on the physical properties of HH containing rare earth metals.


\section{Results and discussion}
\label{sec.res}

\subsection{Electronic band structure}
\label{sec:band}

The bulk electronic band structure along high symmetry directions in the absence and presence of SOC is presented in Fig.~\ref{fig:band}A.
As expected, the introduction of the SOC leads to some band decoupling, which is well visible both in the valence and conduction bands, e.g., at $\Gamma$ point.
Our calculation shows TmPdSb exhibits insulating features with the indirect gap of $\sim 0.17$~eV.
Additionally, we observed $\sim 2$~eV ``gap'' between $\Gamma$-states, which is much larger compared to typical HH insulating compounds with a direct gap.
Hence, the band structure exhibits very untypical character with respect to the typical HH insulators and is similar to the band structure of ScNiBi~\cite{zhang.hou.16}, ScPdBi~\cite{zhang.hou.16,zhang.hou.19}, or Hf$T$Pb ($T=$Ni, Pd, Pt)~\cite{zenati.arbouche.22} reported earlier.

{\it Band order.} As we mentioned in the Introduction just earlier, that the electronic band structure of typical HH compounds resembles that of CdTe/HgTe~\cite{bernevig.hughes.06}.
The bands at the $\Gamma$ point trace back to the $s-$type doublet $\Gamma_{6}$ state, and $p-$type quartet $\Gamma_{8}$ state, whereas, the SOC splits doublet $\Gamma_{7}$ state~\cite{zhu.cheng.12}.
The quartet degeneracy of the $\Gamma_{8}$ state is protected by cubic symmetry.
In such case, the correct order of bands is theoretically reproduced within the meta-GGA mBJ DFT calculations~\cite{zhu.cheng.12,feng.xiao.10,chadov.qi.10}. 
Using this approach, we uncovered that band inversion occurs in TmPdSb.

In the typical case of HH compounds, the realized band inversion is related to that reported in CdTe/HgTe~\cite{zhu.cheng.12,feng.xiao.10,chadov.qi.10}.
In the trivial CdTe, with ``normal'' order of bands at $\Gamma$ point, the $\Gamma_{6}$ state (related to $s$-orbitals) is located above (quartet  degenerated) $\Gamma_{8}$ state.
The band inversion between $\Gamma_{6}$ and $\Gamma_{8}$ states is observed in (topological) HgTe -- in this case, the state $\Gamma_{6}$ is located below $\Gamma_{8}$ state. 
The same band structure behavior is observed in the case of typical HH, 
for example, trivial YPdBi and topological YPtBi~\cite{souza.crivillero.23}.
There, the substitution of Pd in YPdBi by Pt atoms leads to  the ``transition'' from trivial to topological system.
We should mention that this situation is different from that in other topological compounds, where the band inversion can be induced by the turn-on/off spin--orbit coupling~\cite{nachtigal.chong.22}.
In the mentioned case, i.e. YPdBi/YPtBi, the transition is induced by the increasing of the spin--orbit coupling by the atomic substitution~\cite{souza.crivillero.23}.

In the case of TmPdSb, the band inversion (with respect to the ``normal'' order of bands in CdTe) is more complicated.
In typical cases, mentioned above, the inversion occurs between two bands ($\Gamma_{6}$ and $\Gamma_{8}$) closest to the Fermi level.
For TmPdSb, the orbital-projection band structure (see Fig.~\ref{fig:band_inv}(b) in the Supporting Information) uncovers the band inversion in the conduction band at $\Gamma$ point.
Namely, the $\Gamma_{6}$ state is not the first (like in trivial compounds), but the fourth band above the Fermi level. 
In context of the previously reported topological HH, this is novel mechanism of the band inversion.

{\it Role of strain.}
The topological properties can be also modified by the external strain. 
In practice, external strain leads to the lattice parameter modification, what affects the electronic band structure.
Indeed, sometimes strain engineering leads  to the realization of topological states in the HH~\cite{xiao.yao.10,barman.alam.18}.
A similar effect can be obtained by the lattice parameter modification.
This has recently been reported in the case of this film of YPdBi~\cite{palin.adadon.23}, which contrary to YPtBi, possesses trivial bulk features~\cite{souza.crivillero.23}.

In our case, the structural characterization of TmPdSb indicates the lattice parameter of $6.473$~\AA.
We assume that at low temperature ($T\rightarrow\,0$), the lattice parameter should be smaller (corresponding to the positive external hydrostatic pressure).
Similarly, negative external hydrostatic pressure can lead to a larger lattice parameter.
We investigate the band order under these conditions (positive and negative pressure).
Interestingly, we found that the positive pressure led to the shift of $s$-like  $\Gamma_6$-band to higher energies (cf. Fig.~\ref{fig:band_inv}(b) and Fig.~\ref{fig:band_inv}(c) in the Supporting Information), 
which supports the appearance of unusual band order in TmPdSb.
Contrary to this, the negative pressure (increasing of the lattice parameter) leads to the band order well-known form of trivial HH compounds -- i.e. band order observed in CdTe (cf. Fig.~\ref{fig:band_inv}(b) and Fig.~\ref{fig:band_inv}(a) in the Supporting Information).

It is worth mentioning here, that we calculated the topological invariant $\mathbb{Z}_2$ for both the highest occupied band within the valence band, and the $\Gamma_{6}$ band in the conduction band. 
We obtained $\mathbb{Z}_2 = (0;000)$ for the first one, which indicates that the present compound does not exhibit topological insulating properties. 
In contrast, for the band corresponding to the $\Gamma_{6}$ state, $\mathbb{Z}_2 = (1;000)$, indicating a propensity towards topological characteristics.

{\it Surface states.}
We calculated the surface states for different surface orientations (see Fig.~\ref{fig:band}E) of the compound TmPdSb.
The (001)-oriented surface may have TmSb- or Pd-termination. 
Similarly (111) orientation corresponds to the surface with TmPd or SbPd termination.
On the other hand, (011)-oriented surface has only one  termination with all types of atoms within TmPdSb. 
These surface orientations correspond to the projection of the three-dimensional (3D) bulk Brillouin zone on the two-dimensional (2D) surface Brillouin zones, presented in Fig.~\ref{fig:band}F.
Such projection leads to a relatively complex band structure, which is typically a consequence of the projection of several bulk high symmetry points on 2D Brillouin zone (e.g. X and W point on $\bar{\text{M}}$, or $\Gamma$ and X on $\bar{\Gamma}$ for (001) orientation).
Slab-like calculations suggest the realization of several bands of metallic surface states (see Fig.~\ref{fig.slab} in the Supporting Information).
Similar results are obtained by the surface Green function calculations for a semi-infinite system.

First, the surface states are realized over the whole Brillouin zone.
Similar behavior of surface states has earlier been reported in a few semimetallic HH compounds~\cite{liu.yang.16, hosen.dhakal.20,souza.crivillero.23,palin.adadon.23,fang.wu.23}.
Similar surface states realized over the whole Brillouin zone were earlier reported for Bi$_{1-x}$Sb$_{x}$ alloy~\cite{zhu.hofmann.14}, which for $x \geqslant 0.09$ is a topological insulator~\cite{hsieh.qian.08}.
For (001) surface (Fig.~\ref{fig:band}G and~\ref{fig:band}H), such surface states lay between the top of the valence band (located at $\Gamma$ point and projected onto $\bar{\Gamma}$) and the bottom of the conduction band (located at X bulk point and projected onto $\bar{\Gamma}$ and $\bar{\text{M}}$). 
Moreover, the surface states are realized independently for the TmPd or Sb surface termination, and with energies corresponding to the reported band inversion (around $1$~eV above $E_{\rm F}$), which can indicate the topological character 
of TmPdSb (see Fig.~\ref{fig:band}G and Fig.~\ref{fig:band}H).
Similarly, the surface states are observed in the case of (011) and (111) terminations (see Fig.~\ref{fig:band}I and Fig.~\ref{fig:band}J, respectively).
Interestingly, on the (111) surface with TmPd termination, the surface states are not realized.
Additionally, the surface states can be simply distinguished from the bulk states (see also Fig.~\ref{fig.slab} in Supporting Information).

Second, the compounds possessing SS exhibit a non-saturating MR with a linear high-field component and high charge-carrier mobility. 
A few HH compounds also showed such kind of behavior, and it has initially been suggested that it originated from the surface states. 
Nevertheless, it has later been established from the Shubnikov-de Haas oscillations measurements (SdH)~\cite{wosnitza.goll.06, butch.syers.11} that small 3D Fermi surfaces are responsible for such high mobility rather than the SS. 
Our theoretical investigation exhibits the insulating character of TmPdSb (Fig.~\ref{fig:band}A).
However, another topological insulator Bi$_{2}$Se$_{3}$~\cite{pan.vescovo.11,zhu.levy.11}, has been reported to display metallic character due to intrinsic defects~\cite{jia.ji.11,ren.taskin.12}, which shifted its donor level above the bottom of conduction band.
For TmPdSb, the shift of the Fermi level to the bottom of the conduction band (level marked by the pink dashed line on Fig.~\ref{fig:band}A) leads to emerging a small Fermi surface (Fig.~\ref{fig:band}B) in TmPdSb. 
However, we did not observe SdH oscillations in magnetic fields up to 14~T. 
It may be pointed out here, that the magnitude of SdH oscillations depends primarily on effective masses of carriers, but also on level of disorder and number of defects in the crystal.
The disorder in HH compounds is not restricted to antisites or vacancies, but extended to the splitting of the high symmetry Wyckoff positions, as it has been discussed for another  HH compound ScPtBi~\cite{ScPtBi}.

\begin{figure*}
\centering
\includegraphics[width=0.99\textwidth]{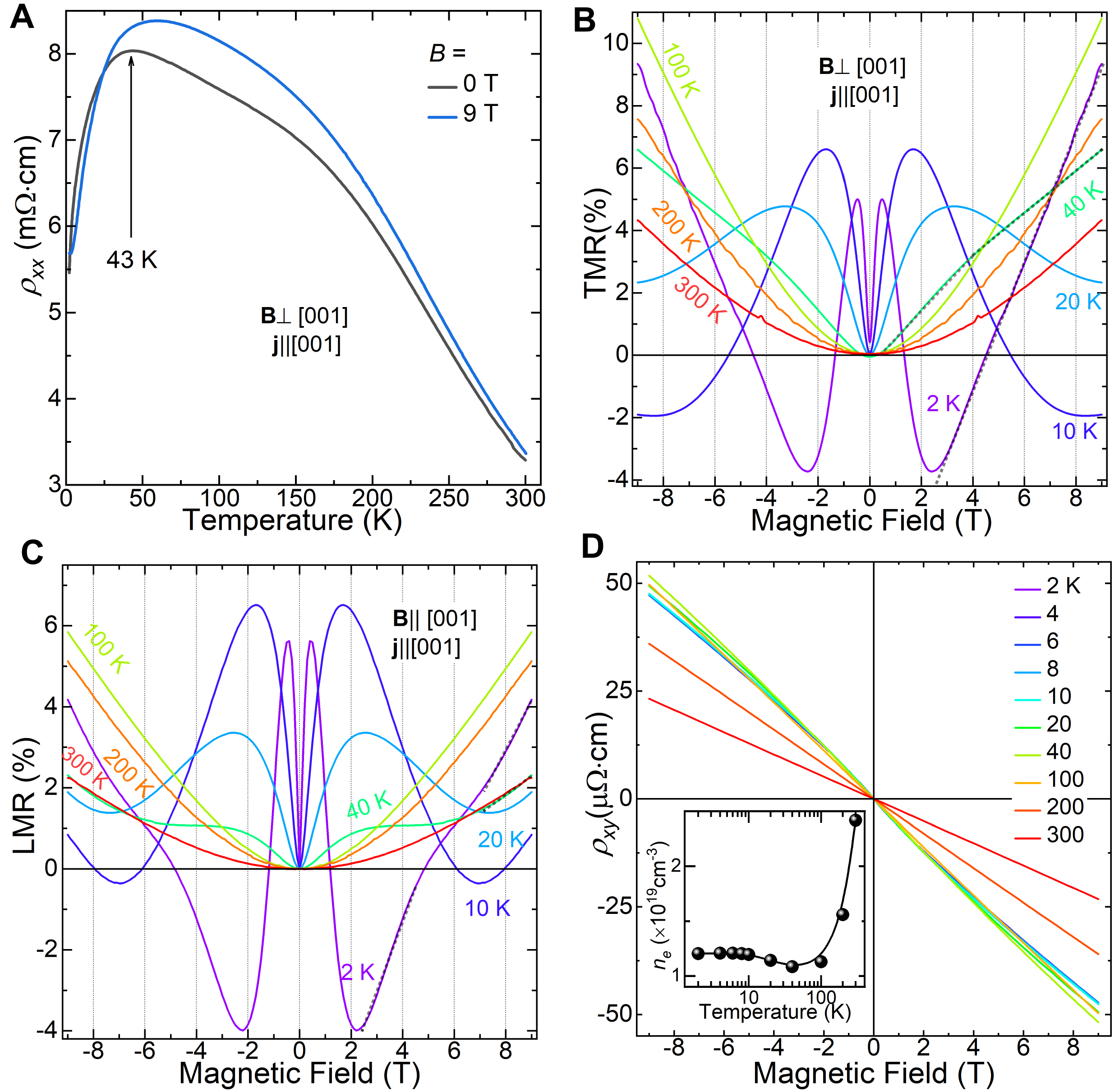}
\caption{
\textbf{Electronic transport.} (\textbf{A}) $\rho_{xx}(T)$ at \textit{B} = 0 and 9~T, respectively, (\textbf{B}) TMR and (\textbf{C}) LMR at different temperatures. Linear portions are highlighted with empty-dotted lines. (\textbf{D}) $\rho_{xy}(B)$ measured at different temperatures, (inset) Electron density ($n_e$) obtained from Hall resistivity in logarithmic scale.
	\label{fig:rho}
 }
\end{figure*}

\subsection{Electronic transport} 
Electrical resistivity of TmPdSb is shown as a function of temperature, $\rho_{xx}(T)$  for $B$ = 0 and 9~T, in Fig.~\ref{fig:rho}A. 
A maximum of $\rho_{xx}(T,B=0)$ data around 45\,K (which persists, but shifts to slightly higher $T$ in a magnetic field  of 9~T) suggests that increase in concentration of thermally generated carriers surmounts carrier scattering on impurities and phonons. 
A similar kind of behavior has been observed in several HH compounds~\cite{nakajima.hu.15}. 
Nevertheless, for TmPdSb we obtained a band gap of 45~meV (Sec.~\ref{sec.sm_2} in Supporting Information). 
Magnetoresistance ($\text{MR} = (\rho_{xx}(B)/\rho_{xx}(0))-1$) was measured in the longitudinal (\textbf{B}$\parallel$\textbf{j}$\parallel$[001]) as well as transverse (\textbf{B}$\bot$\textbf{j}$\parallel$[001]) configurations in the temperature range 2--300~K and in magnetic fields up to 9~T. In both longitudinal magnetoresistance (LMR) and transverse magnetoresistance (TMR), an abrupt increase of MR was observed in low magnetic fields below $\approx$2\,T (Fig.~\ref{fig:rho}B and C), which is typically argued to germinate from WAL. 
We will discuss the WAL behavior in detail, in the next section. 
However, both LMR and TMR show a quadratic ($\sim B^2$) behavior above 40~K. 
It may be noted, that the $\rho_{xx}(T)$ is semiconducting-like at higher temperatures.
Surprisingly, at 40~K, the TMR changes rather linearly with the increase of $B$ in two different regions (shown with black empty-dotted lines in Fig.~\ref{fig:rho}B). On the other hand, the LMR at the same temperature is quite complex; it increases subtly influenced by the WAL and, after a slight decrease, starts behaving quadratically (Fig.~\ref{fig:rho}B). 
Interestingly, at 2~K, the TMR varies almost linearly, when $B> 3$\,T. 
On contrary, the LMR at the same temperature also shows linear behavior with two different slopes in the magnetic field regions 2--5~T and 7--9 T. 
In the high magnetic field, the slope is much steeper. The linear MR has earlier been proposed by Abrikosov to have a quantum origin~\cite{Abrikosov}. 
According to his model, the MR at a high magnetic field and low temperature occurs in systems with linear energy-momentum band dispersion by the population of the lowest Landau level when the thermal energy cannot overcome the Landau level spacing~\cite{Abrikosov,Huang}. It is completely unlikely that at 40\,K and in low-fields TmPdSb is in quantum limit. Therefore Abrikosov model seems inadequate for this compound.
Another model for the linear MR observed even for low carrier density and weak magnetic fields, has been developed for Ag$_{2+\delta}$Se and Ag$_{2+\delta}$Te~\cite{Parish2003}. That model relied on inhomogeneous carrier mobility caused by disorder seems more adequate to TmPdSb, one of HH compounds notorious for inherent structural disorder.
A similar kind of linear MR was observed in several topological compounds, viz., Bi$_{2}$Te$_{3}$, Bi$_{2}$Se$_{3}$, Sb$_{2}$SeTe$_{2}$, MnBi$_{2}$Te$_{4}$ etc. \cite{Huang, Bi2Te3, Tang.Hao.11, Lei} The linear MR of Bi$_{2}$Se$_{3}$ nanoribbons was suggested to result from purely classical mechanism in a two-dimensional system with linear energy dispersion, appearing when Landau levels overlap, even in weak magnetic fields~\cite{Wang}. 
In the case of MnBi$_{2}$Te$_{4}$, a linear MR observed up to a very high temperatures (but only in fields above 5~T) was attributed to SS evidenced by ARPES study~\cite{Lei,Estyunin}. Following our theoretical prediction of SS (Sec. \textbf{Electronic band structure}), a contribution to linear MR related to them can be seriously considered in the case of TmPdSb.

Moreover, in both LMR and TMR, a negative MR (nMR) was observed at 2 K. 
nMR is primarily attributed to two reasons: chiral magnetic anomaly in Weyl semimetals or reduction of magnetic-disorder scattering~\cite{Binghai}. 
However, our band structure calculations (Sec. \textbf{Electronic band structure}) did not show the presence of Weyl states near the Fermi level. On the other hand, the nMR may originate from the magnetic-disorder scattering, i.e., de Gennes-Friedel mechanism~\cite{dgf}. 
We will attempt to account for the contribution from the magnetic-disorder scattering after understanding the contribution from the WAL.

The Hall resistivity ($\rho_{xy}$) at different temperatures (shown in Fig.~\ref{fig:rho}D; antisymmetrized data), reveals a dominance of electron-type carriers all over the measured temperature window. The carrier concentration at different temperatures was estimated from the linear fit of the $\rho_{xy}(B)$ data using the relation $R_{\rm H}= 1/ne$, where, $R_{\rm H}$ is the Hall coefficient.

\begin{figure}[h!]
\centering
\includegraphics[width=\columnwidth]{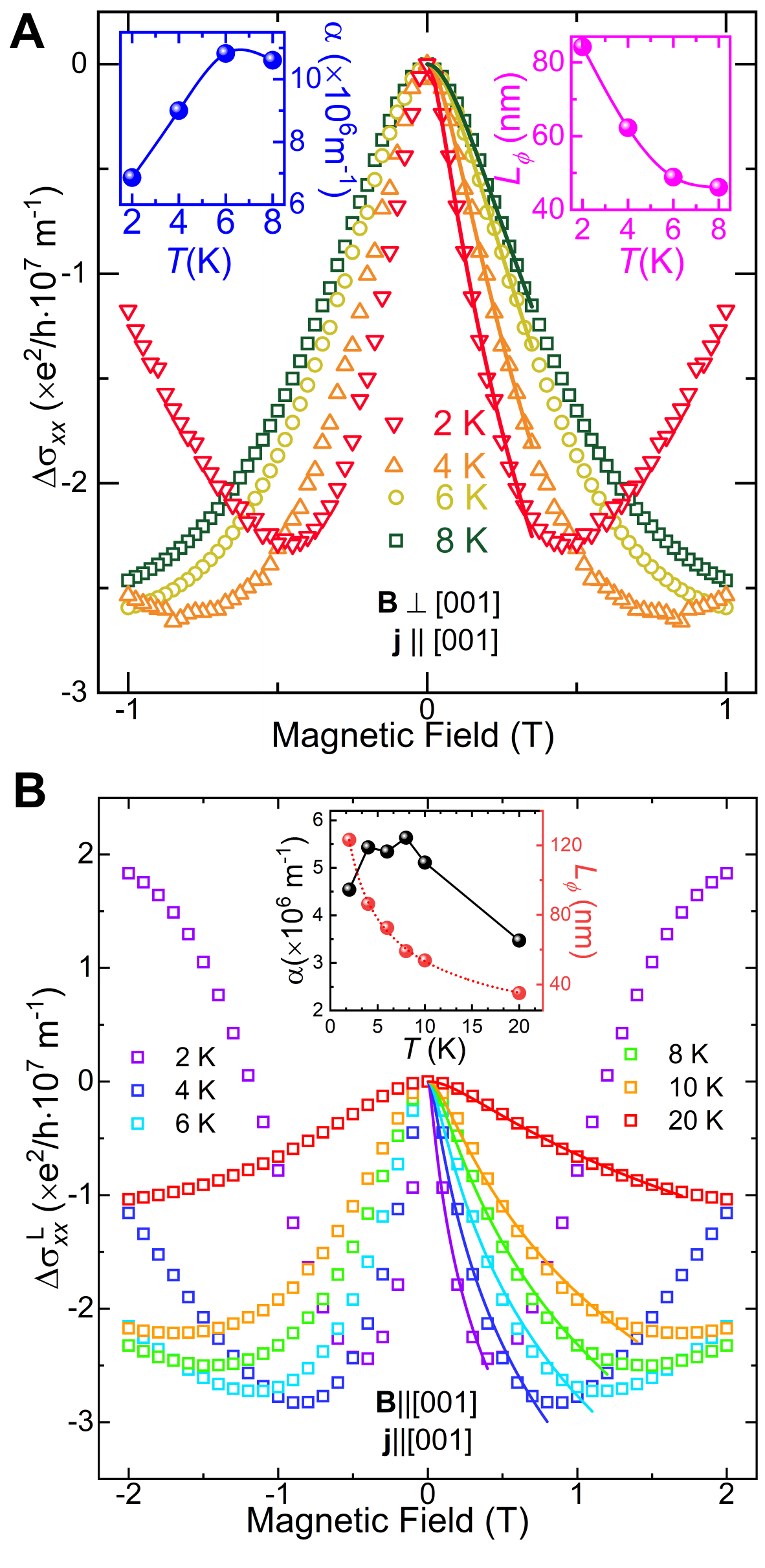}
\caption{
    \textbf{Weak antilocalization.} (\textbf{A}) $\Delta \sigma _{xx}(B)$ and (\textbf{B}) $\Delta \sigma _{xx}^{L}(B)$ in reduced unit at different temperatures. Solid lines with identical colors to the scattered points correlate to the HLN fit, insets show the variation of $\alpha$ and $L_{\phi}$ with temperature.
\label{fig:WAL}
 }
\end{figure}

\begin{figure*} [ht]
\centering
\includegraphics[width=\textwidth]{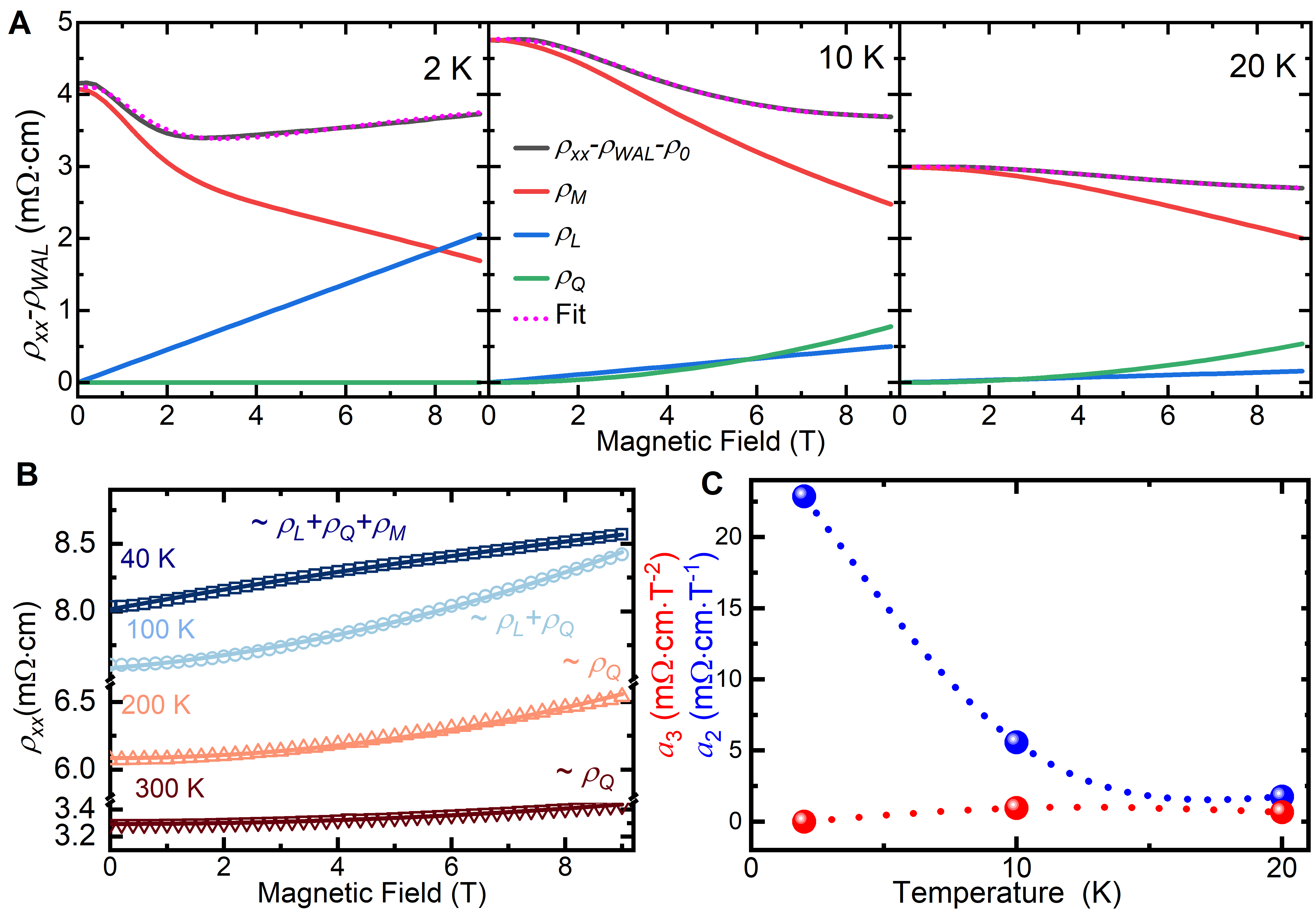}
\caption{
    \textbf{Decomposed $\rho_{xx}(B)$ containing contribution from WAL, magnetic, linear and quadratic components} at $T$ = (\textbf{A}) $2$~K, $10$~K, $20$~K and (\textbf{B}) $40$, $100$, $200$ \& $300$~K. Solid lines with corresponding colors represent the fitting results. (\textbf{C}) The coefficients from linear ($a_2$) and quadratic ($a_3$) $B$ dependence as a function of temperature. Dotted lines are guides to the eye.
    \label{fig:decomposed}
}
\end{figure*}

\subsection{Weak antilocalization}

Components of the conductivity tensor of the compound TmPdSb were calculated from the relations $\sigma_{xx}=\rho_{xx}/(\rho^2_{xx}+\rho^2_{xy})$ and $\sigma_{xy}=\rho_{xy}/(\rho^2_{xx}+\rho^2_{xy})$, and shown in Fig.~\ref{fig:sigma} of the Supporting Information. 
However, for \textbf{j}$\parallel $\textbf{B}, in the absence of Lorentz force, we got the longitudinal conductivity ($\sigma_{xx}^{L}$) as a reciprocal of $\rho_{xx}^{L}$. 
The abrupt decrease of conductivity upon application of weak magnetic field, has earlier been explained within the renormalization group theory framework by Hikami, Larkin and Nagaoka (HLN)~\cite{HLN}. 
Although initially developed for the two-dimensional (2D) conducting sheets, later that approach was used to a number of bulk crystals with strong SOC. According to the model, the change in conductivity ($\Delta \sigma_{xx}$), can be described as:
\begin{eqnarray} 
\nonumber \Delta\sigma_{xx}(B) = \frac{\alpha e^{2}}{2\pi^{2}\hbar}\left[\psi  \left(\frac{1}{2}+\frac{\hbar}{4eL_\phi^2B}\right)+\ln \left( \frac{\hbar}{4eL_\phi^2B}\right )  \right ] \\
\label{eqn:wal}
\end{eqnarray}
\noindent where $\psi$ and $L_{\phi}$ are the digamma function and the phase coherence length, respectively.  
The other symbols carry their usual meaning. We fitted our experimental data with this model for both LMR and TMR. 
Fitting of the Eq.~(\ref{eqn:wal}) yields $\alpha \sim$10$^6$ m$^{-1}$ for both TMR and LMR with a subtle change with temperature (see insets to Fig.~\ref{fig:WAL}A,B).
The value of $\alpha$ is restricted to $\frac{1}{2}$ by the HLN model for a purely 2D sheet, although, high values of the $\alpha$ were reported in several bulk crystals and avowed to be germinated from the enormous number of bulk conduction channels~\cite{CaCuSb, LuPdBi,YPtBi,Bi2Te3}. 
Actually, the designation of the number $\frac{1}{2}$ per conduction sheet simply converts $\alpha$ to the number of conduction channels~\cite{Bi2Te3,CaCuSb}. 
In this context, we should mention that several well-known topological compounds, possessing TSS, e.g.,  Bi$_2$Te$_3$, Bi$_2$Se$_3$, YPtBi, show similar large values of $\alpha$~\cite{YPtBi,Bi2Te3,Checkelsky.2009}. 
A large number of conduction channels has been argued to have novel
technological importance. This can be useful in topological catalysis or in the topoelectronic
devices~\cite{Lee2018,Xie2022,Li2020,Meinzer2017}.
Nevertheless, the variation of $L_{\phi}$ with temperature can be associated with the relation:
\begin{equation} 
\frac{1}{L_\phi^2(T)} = \frac{1}{L_\phi^2(0)}+A_{ee}T+A_{ep}T^2 
\label{eqn:L_phi}
\end{equation}
where $A_{ee}$ \& $A_{ep}$ are the coefficient of the electron-electron and electron-phonon scatterings, respectively, and $L_{\phi}(0)$ corresponds to the phase coherence length at absolute zero temperature. 
While estimating the value of $L_{\phi}(0)$ for transverse and longitudinal direction, we observed an interesting phenomenon. The presence of electron-phonon scattering is negligible in the TMR up to 8~K, whereas the contribution of said interaction is although present in LMR, but its strength is weaker (Table~\ref{tab:WAL}). 
More interestingly, although, the value of the characteristic length 
($\sim$ 86 nm for transverse, and $\sim$120 nm for longitudinal direction, respectively) at 2~K is comparable to other iso-structural compounds, like, LuPdBi, YPtBi and other compounds like, CaCuSb, Bi$_2$Te$_3$ etc., \cite{CaCuSb, LuPdBi,YPtBi,Bi2Te3} the value of $L_{\phi}(0)$ is beyond comparison with the above-mentioned compounds rather comparable to the thin films of Bi$_2$Te$_3$ and Bi$_2$Se$_3$\cite{roy,Minhao}. 
For a 2D material, it was argued that the phase coherence length is closely equal to, or even higher than the thickness of the sample. 
However, the origin of the large value of $L_{\phi}$ was argued to originate only from the surface conductivity. 
We also wish to mention here, that normally, $L_{\phi}$ tends to saturate while decreasing the temperature, which does not happen for our data. Moreover, it has been argued that $L_{\phi}$ is dependent on temperature as a power function with an exponent close to $-0.5$ in the case of 2D WAL, and close to $-0.75$ for a 3D WAL behavior~\cite{bao2012weak, altshuler1982effects}. For our results, both TMR and LMR, we fitted the temperature dependence of $L_{\phi}$, as shown in Fig. \ref{fig:HLN_2D} of the Supporting Information. The exponent values clearly indicate that the WAL has 2D character, and therefore is related to the surface.

\begin{table}[h]
\caption{Coefficients of Eq.~(\ref{eqn:L_phi}) obtained for both TMR and LMR.}
\label{tab:WAL}
\begin{tabular}{ll|ll}
\multicolumn{2}{c}{TMR} & \multicolumn{2}{c}{LMR} \\ 
\hline
$A_{ee}$ & 6.08$\times$ 10$^{-5}$ nm$^{-2}$K$^{-1}$ & $A_{ee}$ & 2.92$\times$ 10$^{-5}$ nm$^{-2}$K$^{-1}$ \\ 
\hline
$A_{ep}$ & 0 & $A_{ep}$ & 5.352$\times$ 10$^{-7}$ nm$^{-2}$K$^{-2}$ \\ 
\hline
$L_{\phi}(0)$ & 227 nm & $L_{\phi}(0)$ & 434 nm                 
\end{tabular}
\end{table}

In stronger fields and at $T<40$~K the behavior of magnetoresistance is very complex and seems to arise due to several mechanisms. We analyze it in the following section.


\subsection{Decomposition of magnetoresistance}

To unravel the complex magnetoresistivity in higher magnetic fields, we subtracted its component arising from WAL ($\rho_{\mathrm{WAL}}$). We simply extrapolated Eq.~(\ref{fig:WAL}) to the high field region by fixing the coefficients obtained while fitting in the low-field region followed by a reverse conversion of $\sigma_{xx}$ to $\rho_{xx}$ with the relation, $\rho_{xx}= \sigma_{xx}/(\sigma^2_{xx}+\sigma^2_{xy})$. 
\begin{align}
\label{eqn:decompose}
\rho_{xx} -\rho_{\mathrm{WAL}} &= \rho_{0}+\rho_{M}+\rho_{L}+\rho_{Q}\\
\rho_{M}&= a_1\left(1- \frac{M^2(B)}{M_S^2}\right)\\
\rho_{L}&= a_2B\\
\rho_{Q} &= a_3B^2
\end{align}
where $\rho_{0}$ is the residual resistivity at $B= 0$ and $\rho_{mag}$ is the  magnetic contribution originating from the magnetic spin scattering. 
$a_2$ and $a_3$ are the coefficients from the linear and quadratic contribution, respectively. 
For extracting the magnetic contribution, we used the $M(B)$ (which we measured in field up to 7 T) at different temperatures and extrapolated the values up to 9 T using the relation $M(B)= M_S(T)\tanh \mu B+\chi_0 B$ (cf. inset of Fig.~\ref{fig:M} in the Supporting Information), where $\chi_0$ is used to account for the contribution of the paramagnetic moment. 
We also assumed the saturation magnetic moment ($M_S$) of Tm$^{3+}$ as 7 $\mu_{\rm B}$. Magnetic field dependence of the particular components of magnetoresistance of the compound at 2, 10 and 20 K are plotted in Fig.~\ref{fig:decomposed}A.
As no WAL was observed beyond 20 K, the resistivity as a function of magnetic field at 40, 100, 200 and 300 K  are  shown in Fig~\ref{fig:decomposed}B. 
we also plotted the variation of coefficients $a_2$ and $a_3$ in Fig~\ref{fig:decomposed}C. 
The quadratic contribution may be explained with Drude model. 
As the compound is an insulator (Sec. \textbf{Electronic structure}), with band gap $\sim 0.09$~eV from the experimental evidence and $\sim$0.17\,eV from the theoretical calculation, the linear contribution may originate from the SS predicted by our theoretical calculations.
For ZrSiS, while explaining the complex angular dependence of MR, a similar kind of behavior consisting of both linear and quadratic components has been observed for MR measured for different angles of applied $B$. 
It was argued that the variation of those components with $B$ arises due to the contribution of the projection of bulk Fermi surface as well as of Fermi contour (in 2D)~\cite{ZrSiS_Sci}. 
In the case of TmPdSb, the linear contribution decreases with increasing temperature, which additionally supports the role of topologically non-trivial surface states.  


\section{Summary}
\label{sec.sum}

We studied the electronic properties of the ternary TmPdSb half-Heusler compound.
Theoretical investigations of the band structure uncovered unusual inversion of band order within the conduction band.
Additionally, the metallic surface states were found over the whole Brillouin zone.
We found them using slab-like calculations, as well as the surface Green function calculations for the semi-infinite system.
Typically, topological insulators (TIs) are distinguished by the value of $\mathbb{Z}_2$ invariants. 
Determining whether the surface bands are topologically trivial or non-trivial can distinctly elucidate the material's nature. 
However, it's important to note that the existence of metallic surface states doesn't necessarily indicate the material's topological nontriviality. 
From the experimental point of view, an attempt to prepare single crystals of even better quality and characterize possible quantum oscillations due to SS may be envisaged. 
There are also recent reports on interesting modifications of optical properties of materials due to their topologically nontrivial nature~\cite{Orenstein}, and low-field differential magnetometry providing insight into the properties of SS in topological systems~\cite{Zhao}. 
The Fermi level can be tuned i.e. by electron irradiation~\cite{Ishihara}, which in the case of SS shown in Fig.~\ref{fig:band}G-J could lead to strong modifications of their contribution to magnetotransport.
Nevertheless, correlating the topological status of a material with the unusual band order, like the present case, needs some novel theoretical description, which may germinate a completely new field for future research.

The existence of such conducting surface states was also reflected in the experimental results.
The insulating nature of the compound at temperatures above $250$~K was confirmed experimentally by the electrical resistivity measurement.
The TMR and LMR at low temperatures exhibit a $B$-linear dependence, whereas, at high temperatures, MR shows a usual $B$-quadratic dependence. 
Although TmPdSb remains paramagnetic down to $2$~K, a contribution from magnetic scattering was observed below 40~K.
Large values of $\alpha$ ($\sim 10^{6}$~m$^{-1}$), as well as of coherence length $L_{\phi}(0)$ ($\sim$ 227 nm for TMR and $\sim$ 434 nm for LMR) were derived from analysis of WAL.
Such relatively large values of $L_{\phi}(0)$ are more typical to the 2D topological sheets only, and indicate realization of the surface states. 
Similarly, a large value of $\alpha$ strongly suggests a huge number of conducting channels, which also can be related to the metallic surface states.
Moreover, the decomposition of the magnetoresistance below 40~K shows a linear $B$-dependence, and the magnitude of the linear contribution decreases with increasing temperature.


\section{Methods}
\label{sec.meth}

\subsection{Experimental}

High-quality single crystals of TmPdSb were grown using the Bi flux. At least 99.99\% pure Tm, Pd, Sb, and Bi were placed in an alumina crucible with a molar ratio of 1:1:1:30. The crucible was sealed in an evacuated quartz ampule and placed in a furnace for heating. The temperature of the furnace was increased at a rate of 20$^{\circ}$C/h up to 1150$^{\circ}$C and kept at that temperature for 24 h for homogenization. Then the furnace was allowed to cool at a very slow rate of 2$^{\circ}$C/h down to 800$^{\circ}$C followed by a natural cooling down to room temperature. To separate the crystals from the melt, the ampule was centrifuged at 350$^{\circ}$C. The stoichiometry of the obtained single crystals was checked using a scanning electron microscope (FEI) equipped with an EDS spectrometer (Genesis XM4). The crystals were found homogeneous and single-phase. The powder X-ray diffraction (XRD) study of a few finely crushed crystals was done in a PANanalytical X’pert Pro diffractometer with Cu $K_{\alpha}$ radiation. The refinement of the XRD pattern was carried out using the Rietveld method and {\sc FullProf} software package. The quality of the crystals was checked using the back-scattered Laue diffraction method (using Laue-COS, Proto Manufacturing). Magnetic susceptibility of the compound was measured in a SQUID magnetometer (MPMS-XL, Quantum Design), whereas the magnetic isotherms were recorded in a physical property measurement system (PPMS-14, Quantum Design) with a vibrating sample magnetometer head. The magnetotransport properties of the compound were measured on a bar-shaped (with cross-section $192\times 115~\mu$m$^2$)  piece cut from the oriented crystal. For preparing electrical contact, we used a silver wire with a diameter 20\,$\mu$m glued to the sample with a silver paste. 
The distance between the voltage contacts was 269\,$\mu$m. We symmetrized the resistivity ($\rho_{xx}(B)$) and anti-symmetrized the Hall resistivity ($\rho_{xy}(B)$) with respect to the applied magnetic, so as to eliminate effects of contact misalignment. Structural and magnetic characterization of the compound is presented in the Supporting Information.

\subsection{Computational details}

First-principles (DFT) calculations were performed using the projector augmented-wave (PAW) potentials~\cite{blochl.94} implemented in the Vienna \emph{ab initio} Simulation Package ({\sc Vasp}) code~\cite{kresse.hafner.94,kresse.furthmuller.96,kresse.joubert.99}.
Calculations were made within the meta-generalized gradient approximation (GGA) under the modified Becke-Johnson (mBJ) pseudopotential~\cite{tran.blaha.09}, which correctly allows the study of the topological properties of the HH compounds~\cite{feng.xiao.10,alsawai.lin.10}.
In calculations the $f$ electrons were treated as core states.
The valence electron configuration $4d^{9}5s^{1}$ for Pd, $6s^{2}5p^{6}5d^{1}$ for Tm, and $5s^{2}5p^{3}$ for Sb were taken.
The calculations were performed with the energy cut-off set to $350$~eV.

In calculation, we used the experimental lattice parameters.
The compound crystallized in cubic structure $F\bar{4}3m$ (space group No.~216) with lattice constant $a = 6.473$~\AA~ for a conventional cell containing four formula units.
The atoms' positions are defined by high symmetry Wyckoff positions: $4a$ (0,0,0) for Tm, $4b$ (1/2,1/2,1/2) for Pd, and $4c$ (1/4,1/4,1/4) for Sb.

The topological properties, as well as the electronic surface states, were studied using the tight-binding model in the maximally localized Wannier orbitals basis~\cite{marzari.vanderbilt.97,souza.marzari.01}.
This model was constructed from exact DFT calculations in a primitive unit cell (containing one formula unit), with $10 \times 10 \times 10$ $\Gamma$-centered {\bf k}--point grid, using the {\sc Wannier90} software~\cite{pizzi.vitale.20}.
We started from initial $s$, $p$, and $d$ orbitals on each atom, which allowed to construct tight-binding model taking into account 27 orbitals and 54 bands.
During calculations, the $f$ electrons of Tm were treated as core states.
The electronic surface states were calculated using the surface Green's function technique for a semi-infinite system~\cite{sancho.sancho.85}, implemented in {\sc WannierTools}~\cite{wu.zhang.18}.

\section*{Acknowledgments}

\noindent Authors thank Ms. Ewa Bukowska and Damian Szymański for XRD and ESD measurements, respectively.
This work was supported by National Science Centre (NCN, Poland) under Projects No. 2021/40/Q/ST5/00066 (S.D., O.P., P.W. and D.K.) and 2021/43/B/ST3/02166 (A.P.).

\section*{Conflict of Interest}

\noindent The authors declare no conflict of interest.

\section*{Author contributions}

\noindent Writing - Original Draft: S.D., P.W. and A.P.
Visualization: S.D. and A.P.
Theoretical part: A.P.
Formal analysis: S.D., P.W., O.P. 
Funding acquisition: A.P. and D.K.  
Conceptualization: A.P., S.D., D.K. and P.W. Writing - Review \& Editing S.D., A.P., K.S., O.P., P.W. and D.K.
All authors contributed to the discussion and analysis of the results of the manuscript. 
Furthermore, All authors have read and agreed to the published version of the manuscript.

\section*{Data Availability Statement}

\noindent The data that support the findings of this study are available from the corresponding author upon reasonable request.

\section*{Keywords}

half-Heusler compound, metallic surface states, electronic properties, magnetoresistance

\bibliography{TmPdSb_ref}

\clearpage
\newpage

\onecolumngrid

\begin{center}
  \textbf{\Large Supporting Information}\\[.2cm]
  \textbf{\large Insulating Half-Heusler TmPdSb with Unusual Band Order and Metallic Surface States}\\[.2cm]
  Shovan Dan$^{1}$, Andrzej Ptok$^{2}$, O. Pavlosiuk$^{1}$, Karan Singh$^{1}$, P. Wi\'{s}niewski$^{1}$, and D. Kaczorowski$^{1}$\\[.2cm]
  {\itshape
  	\mbox{$^{1}$Institute of Low Temperature and Structure Research, Polish Academy of Sciences, Okólna 2, 50-422 Wrocław, Poland}\\
	\mbox{$^{2}$Institute of Nuclear Physics, Polish Academy of Sciences, W. E. Radzikowskiego 152, PL-31342 Kraków, Poland}
  }
(Dated: \today)
\\[1cm]
\end{center}

\setcounter{equation}{0}
\renewcommand{\theequation}{S\arabic{equation}}
\setcounter{figure}{0}
\renewcommand{\thefigure}{S\arabic{figure}}
\setcounter{section}{0}
\renewcommand{\thesection}{S\arabic{section}}
\setcounter{table}{0}
\renewcommand{\thetable}{S\arabic{table}}
\setcounter{page}{1}

\begin{figure}[!h]
\centering
\includegraphics[width=\textwidth]{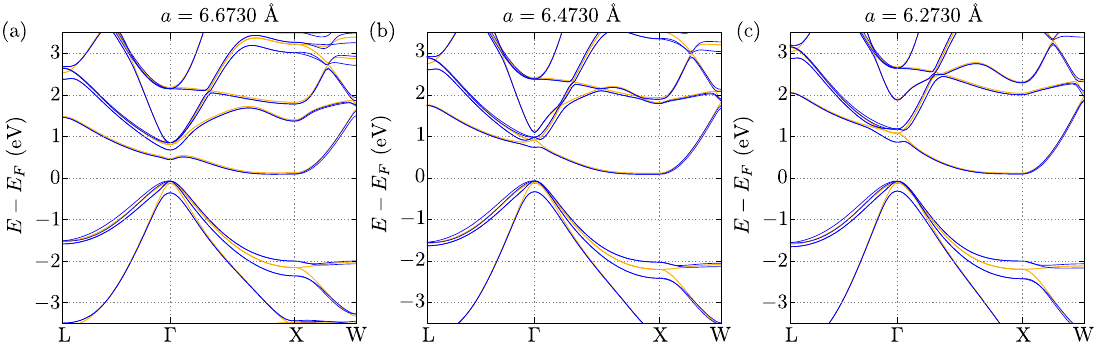}
\caption{
{\bf Role of the spin--orbit coupling} on the electronic band structure of TmPdSb, for different lattice parameters (as labeled).
The orange and blue lines correspond to the absence and to the presence of the spin--orbit coupling, respectively.
Results presented in (b) correspond to the experimentally obtained lattice constant.
The spin--orbit coupling open the gap in some high symmetry point and directions (cf.~orange and blue lines).
\label{fig:band_soc}
}
\end{figure}
\begin{figure}[!hb]
    \centering
    \includegraphics[width=\columnwidth]{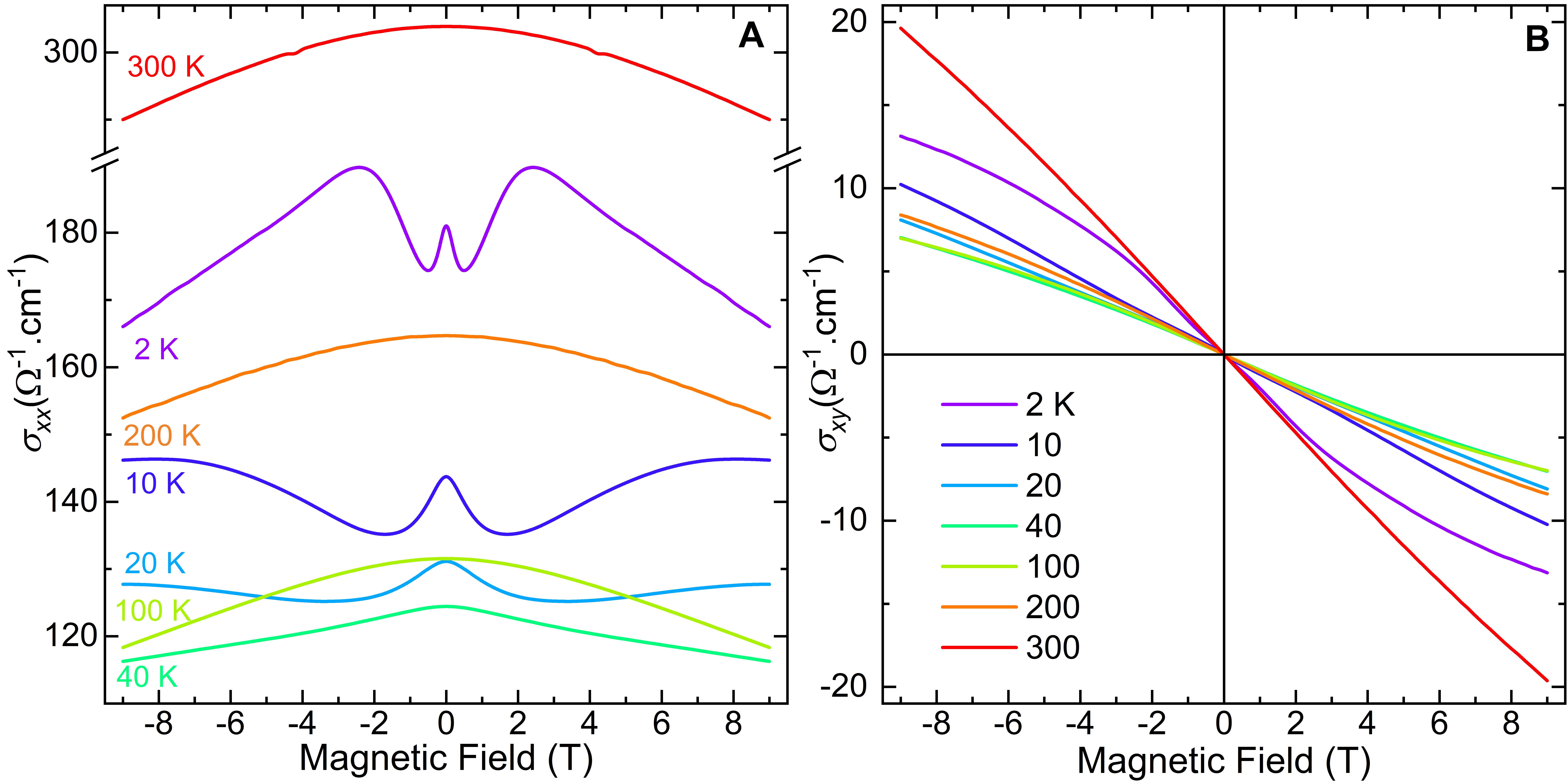}
    \caption{\textbf{Components of the electrical conductivity} of TmPdSb ~(\textbf{A}) longitudinal $\sigma_{xx}(B)$ and (\textbf{B}) Hall $\sigma_{xy}(B)$, measured at different temperatures, in the magnetic field \textbf{B}$\bot$ [001], with current \textbf{j}$\parallel$ [001].}
    \label{fig:sigma}
\end{figure}
\begin{figure}[!hb]
\centering
\includegraphics[width=0.55\columnwidth]{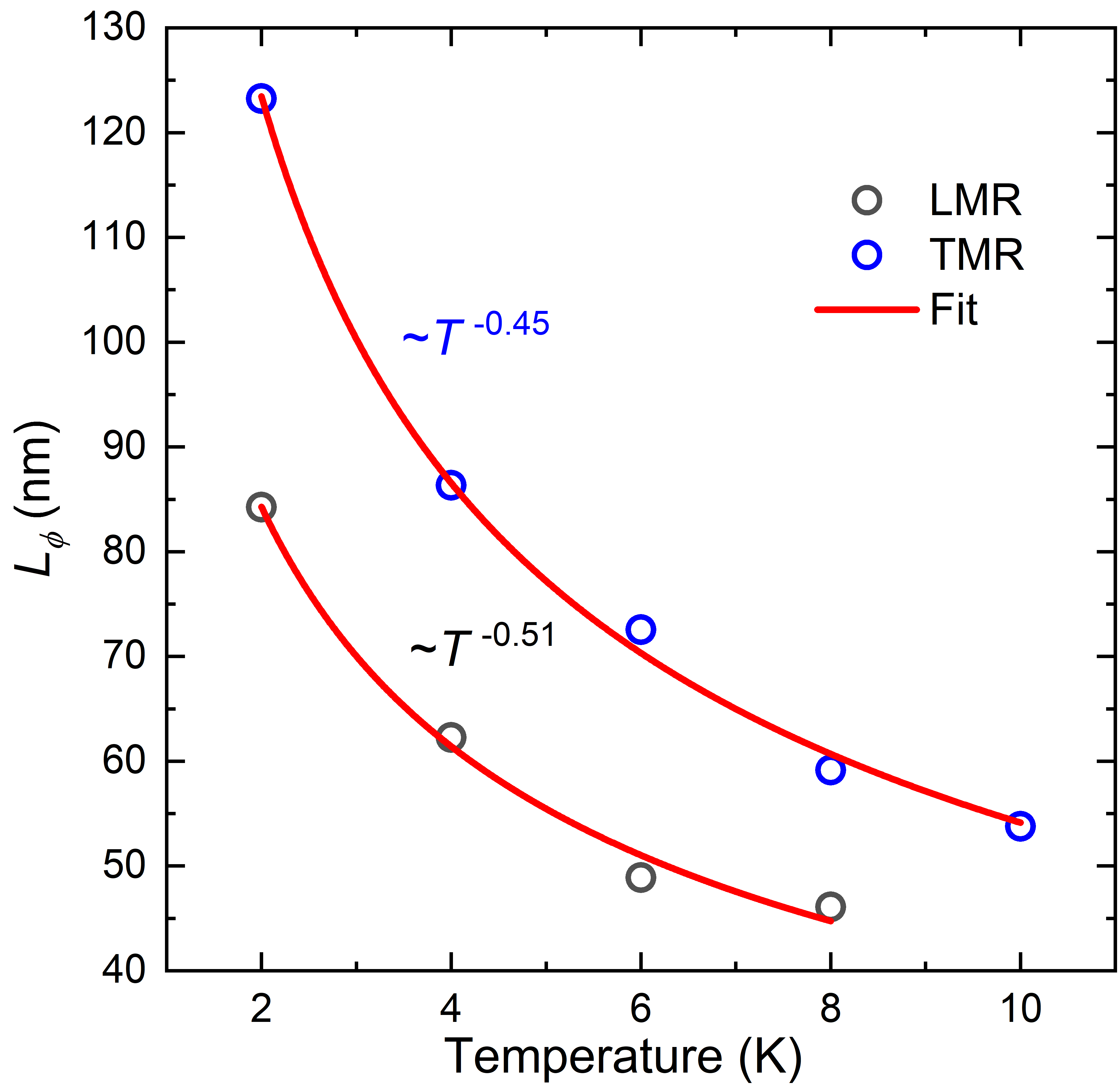}
\caption{
    Variation of the coherence length $L_{\phi}$  (obtained for TMR and LMR) with temperature.
\label{fig:HLN_2D}
}
\end{figure}

In this Supporting Information, we present additional results:
\begin{itemize}
\item Sec.~\ref{sec.band_inv} Band inversion and surface states.
\item ~~~~~~Fig.~\ref{fig:band_inv} Orbital order as a function of lattice parameters.
\item ~~~~~~Fig.~\ref{fig:band_soc} Role of the spin--orbit coupling on the electronic band structure.
\item ~~~~~~Fig.~\ref{fig.slab} Results of the slab-like calculations.
\item Sec.~\ref{sec.sm_2} Band gap. 
\item ~~~~~~Fig.~\ref{fig:bandgap} Band gap.
\item Sec.~\ref{sec.sm_3} Longitudinal and Hall conductivity.
\item ~~~~~~Fig.~\ref{fig:sigma} Components of the electrical conductivity
\item Fig. \ref{fig:HLN_2D} Variation of characteristic length $L_{\phi}$ with temperature.
\item Sec.~\ref{sec.sm_1} Structural and magnetic characterization. 
\item ~~~~~~Fig.~\ref{fig:xrd} Structural characterization.
\item ~~~~~~Fig.~\ref{fig:eds} EDS spectrum. 
\item ~~~~~~Fig.~\ref{fig:M} Magnetic properties.
\item ~~~~~~Fig.~\ref{fig:MH14} High field magnetization isotherm.
\end{itemize}

\newpage

\section{Band inversion and surface states}
\label{sec.band_inv}

\begin{figure}[!h]
\centering
\includegraphics[width=\textwidth]{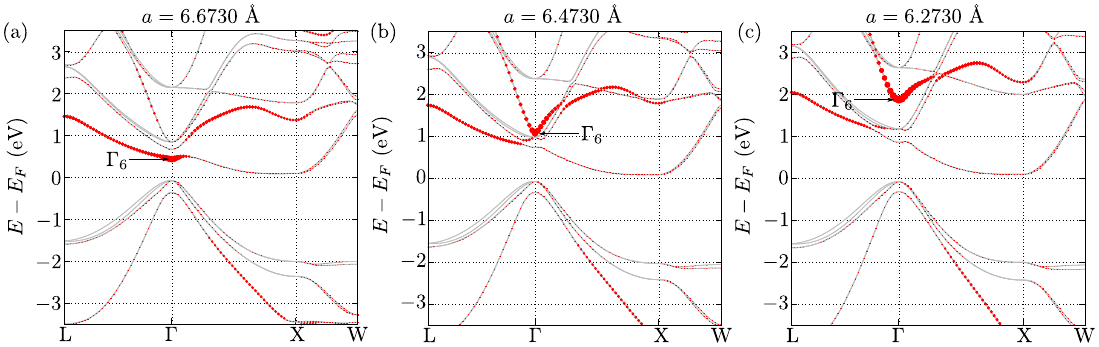}
\caption{
{\bf Realization of the unusual band order} of TmPdSb, for different lattice parameters (as labeled).
Results for the experimental lattice parameter are presented on panel (b).
The size of the red dot represents $s$-orbital contribution.
The band order presented on (a) corresponds to the ``normal'' order (similar to this reported for CdTe).
Similarly, band order on (b) and (c) corresponds to the unusual band inversion (around $1$~eV).
Results in the presence of spin--orbit coupling.
Results presented in (b) correspond to the experimentally obtained lattice constant.
\label{fig:band_inv}
}
\end{figure}

\begin{figure}[!h]
\centering
\includegraphics[width=0.7\textwidth]{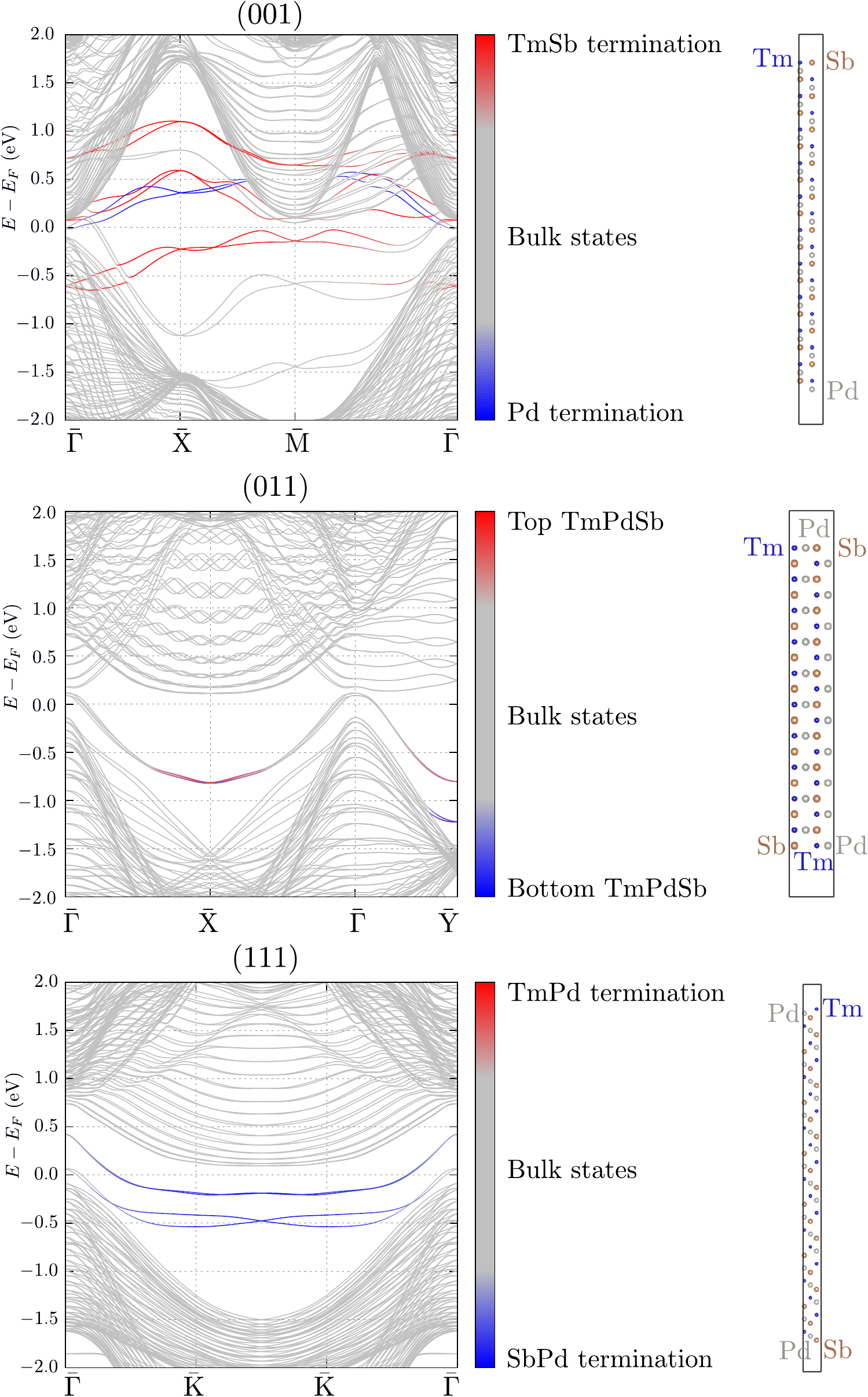}
\caption{
{\bf Theoretically obtained spectra} for the surfaces with different orientations (as labeled).
Results of slab-like calculations.
The colors of lines correspond to the bulk or surface layers (as labeled).
The right panels present corresponding unit cells containing 20 formula units.
Related surface Green function are presented in Fig.~\ref{fig:band} in the main text.
Surface (011) realized for the same termination for top and bottom surfaces.
In the case of (111) surface, there are no surface states corresponding to the top surface with TmPd termination.
\label{fig.slab}
}
\end{figure}

\newpage


\section{Band gap}
\label{sec.sm_2}

\begin{figure}[!h]
    \centering
    \includegraphics[width=0.65\columnwidth]{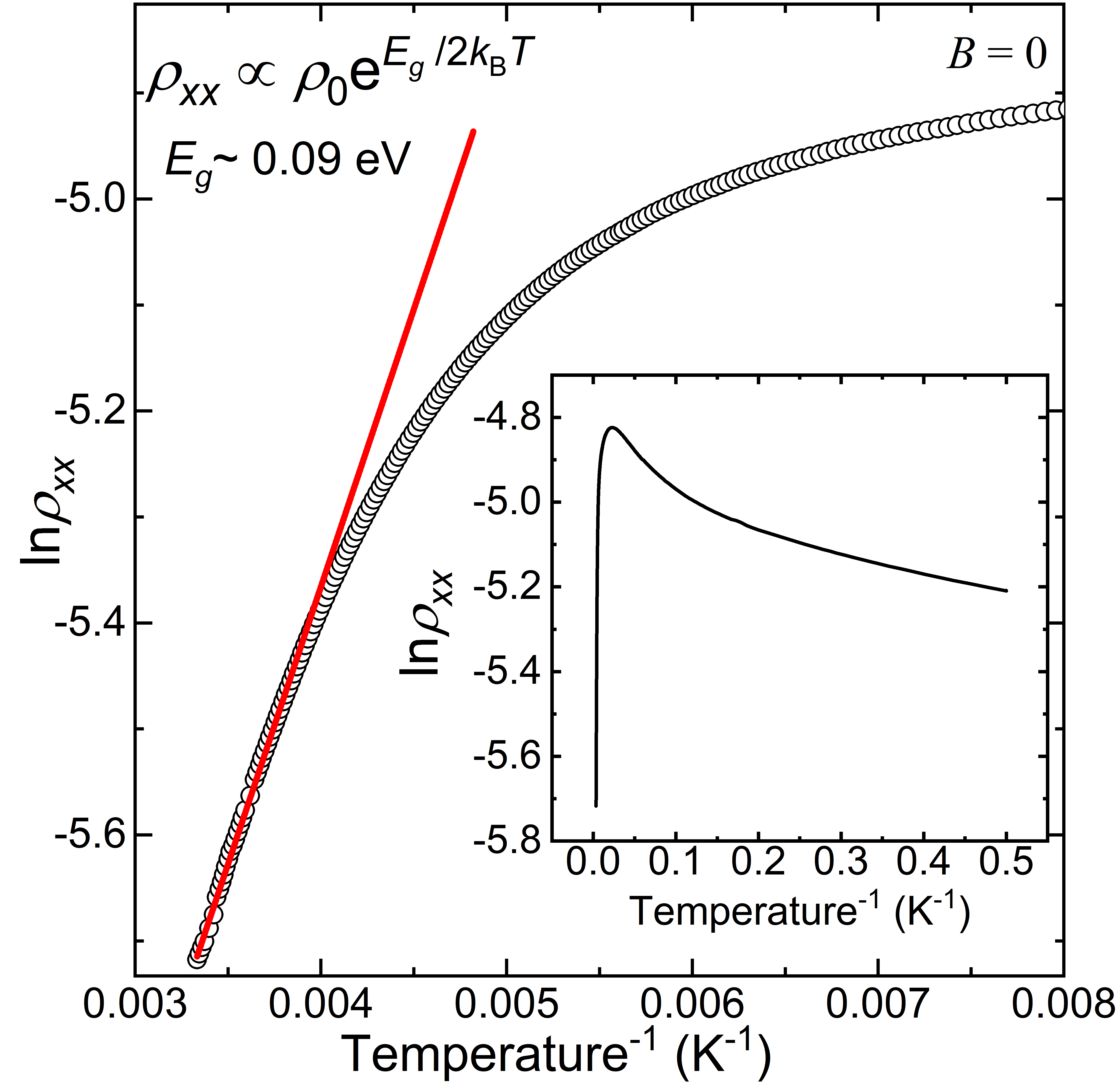}
    \caption{\textbf{Band gap.} $\ln \rho_{xx}$ vs. $1/T$ plot to evaluate band gap. Inset shows data in the entire $2$--$300$~K region.}
    \label{fig:bandgap}
\end{figure}

To evaluate the electronic band gap at room temperature, we plotted $\ln \rho_{xx}(B=0)$ vs. $1/T$, as shown in Fig.~\ref{fig:bandgap}, and fitted a function: $\rho_{xx} \propto\rho_0\exp(E_g/2k_{\rm B}T)$, in $250$--$300$~K range, which yielded a band-gap $E_g=0.09$~eV.

\newpage

\section{Longitudinal and Hall conductivity}
\label{sec.sm_3}

\newpage

\section{Structural and magnetic characterization}
\label{sec.sm_1}

\begin{figure}[!b]
    \centering
    \includegraphics[width=\columnwidth]{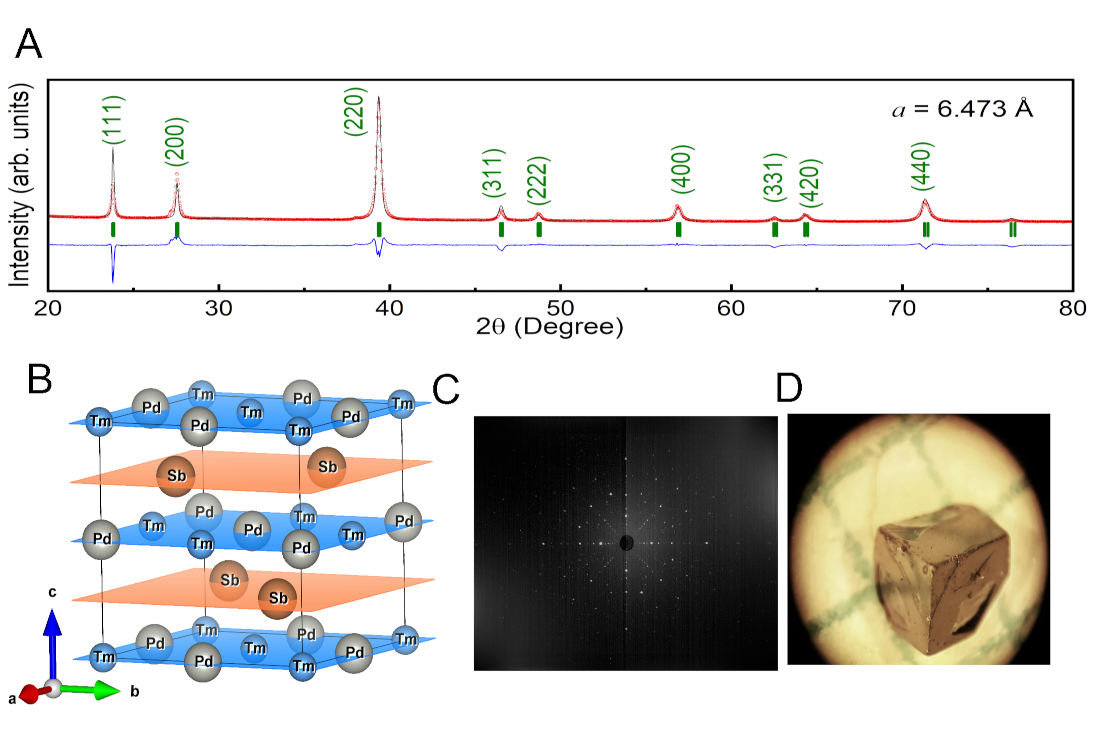}
    \caption{\textbf{Structural Characterization.} (\textbf{A}) Room temperature powder-XRD pattern. The red circles, black line, blue line, and olive bars represent, experimental data, simulated pattern, the difference between experimental and simulated data, and Bragg reflections allowed by the crystal symmetry, respectively. (\textbf{B}) Crystal structure with Tm-Sb and Pd layers, (\textbf{C}) Laue pattern for X-ray beam along [001] direction, and  (\textbf{D}) the photograph of the crystal.}
    \label{fig:xrd}
\end{figure}

\begin{figure}[!pt]
    \centering
    \includegraphics[width=0.7\columnwidth]{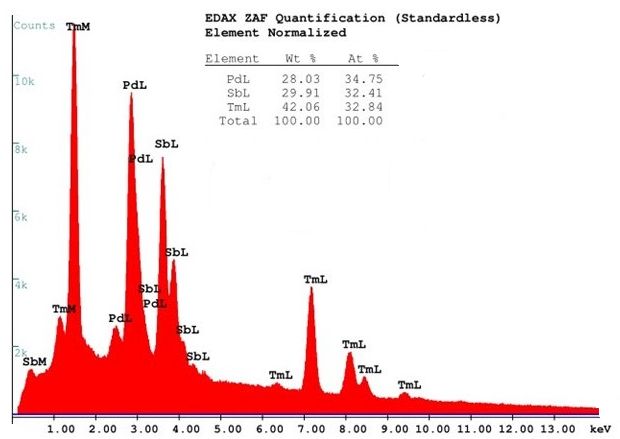}
    \caption{\textbf{EDS spectrum} for the single crystal of TmPdSb, with weight and atomic fractions of elements shown in the inset.}
    \label{fig:eds}
\end{figure}

Fig.~\ref{fig:xrd}(a) shows the room temperature XRD pattern of TmPdSb. The compound crystallizes in the MgAgAs-type crystal structure, with space group $F\bar{4}3m$ (Fig.~\ref{fig:xrd}(b)). 
Back-scattering Laue diffractogram for the (001)-direction, and the photograph of a representative crystal are shown in Figs.~\ref{fig:xrd}(c) and (d), respectively. 
The EDS spectra measured in several points of the crystals avowed the elemental stoichiometry (Fig. \ref{fig:eds}).

\begin{figure}[!pb]
	\centering
	\includegraphics[width=0.6\columnwidth]{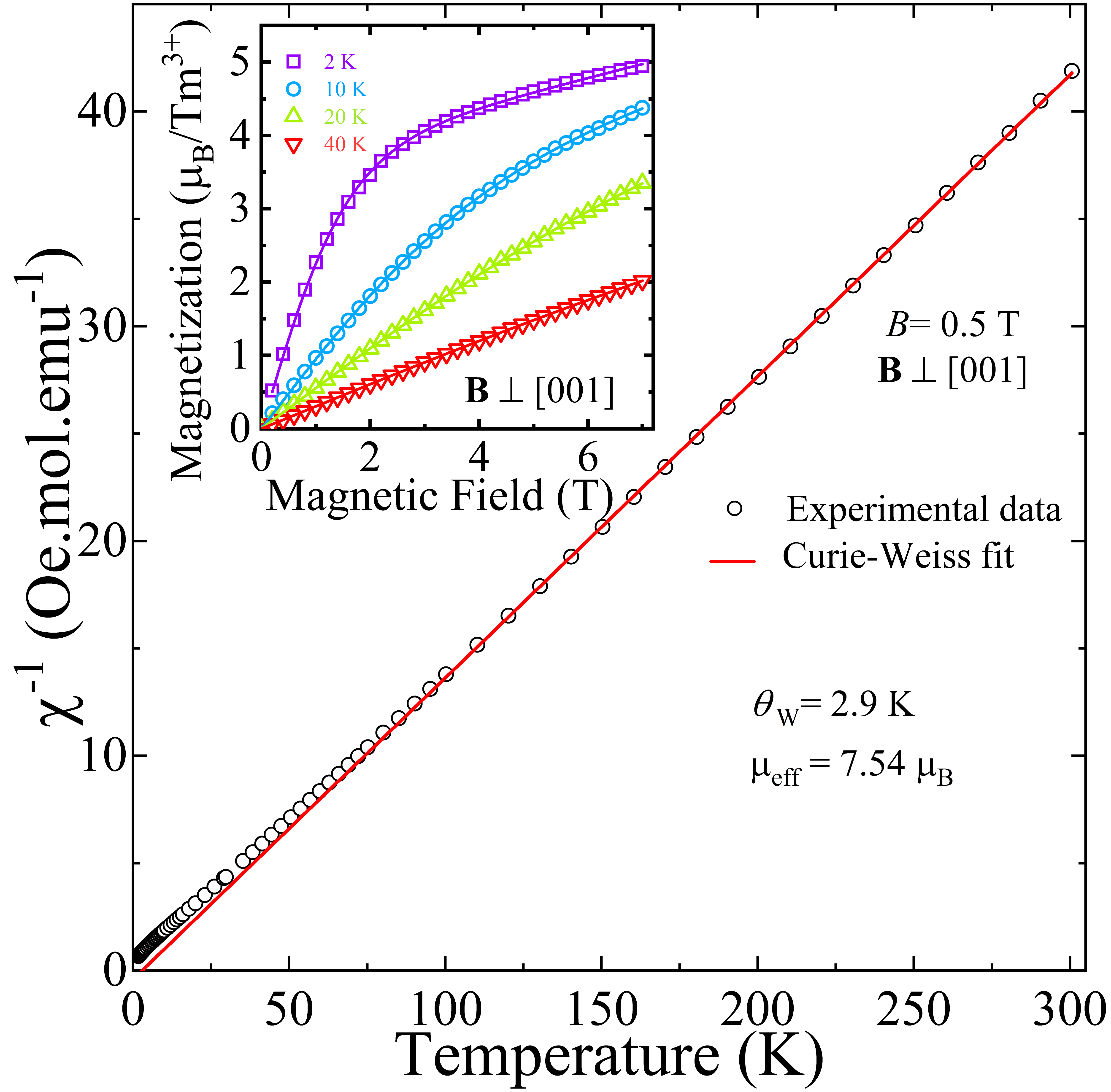}
	\caption{\textbf{Magnetic properties}. Inverse susceptibility with Curie-Weiss fit, measured in $B= 0.5$~T. Inset shows magnetic isotherms of the compound measures at $T= 2$, $10$, $20$, and $40$~K. The solid lines of colors identical to those of the scattered points correspond to fits with $M(B)= M_S\tanh \mu B+ \chi_0 B$ equation (see main text).}
	\label{fig:M}
\end{figure}

\begin{figure}[!t]
    \centering
    \includegraphics[width=0.65\columnwidth]{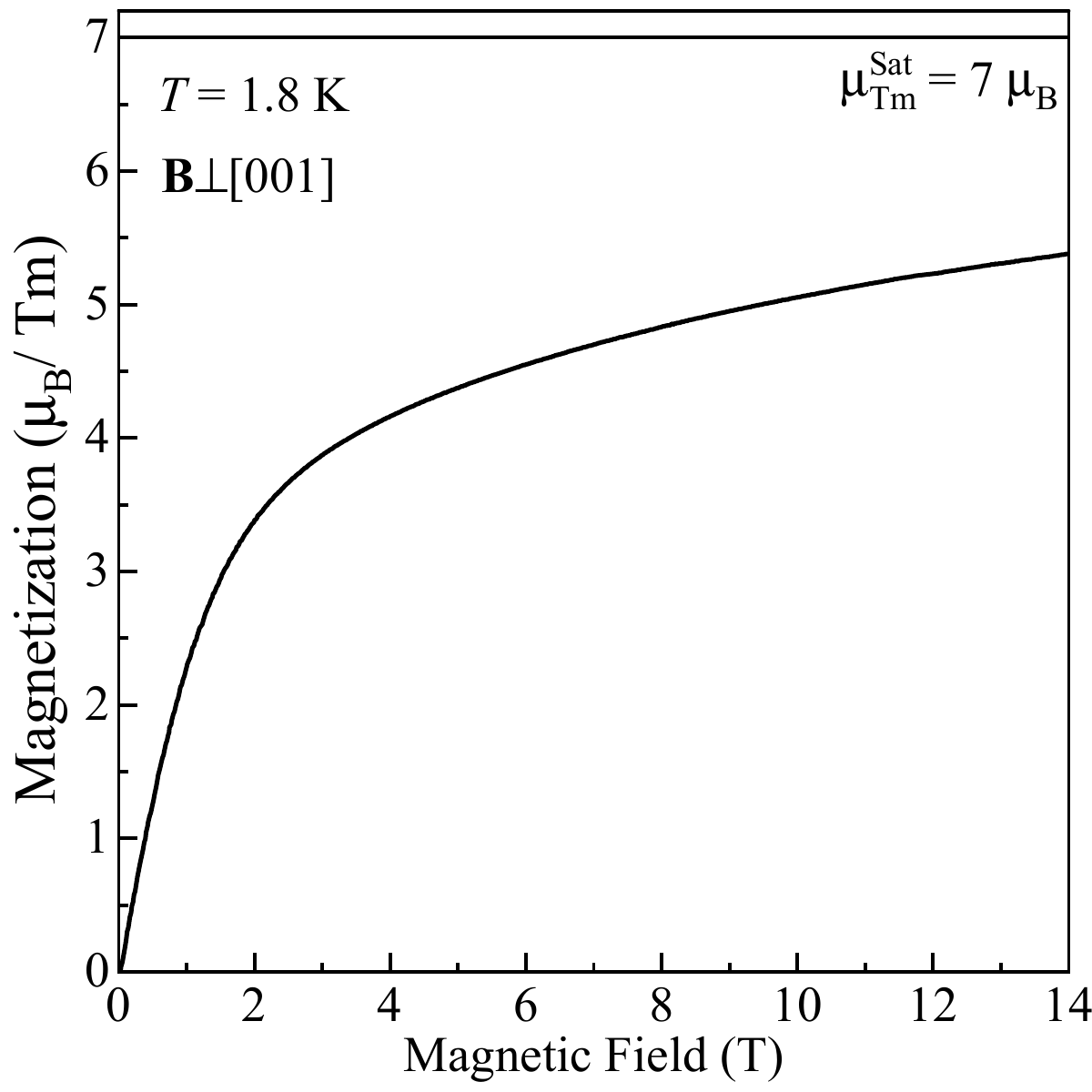}
    \caption{\textbf{High field magnetization isotherm} measured at $T = 1.8$~K in magnetic field up to $14$~T.}
    \label{fig:MH14}
\end{figure}

The magnetic susceptibility ($\chi$) of TmPdSb was measured in the temperature region $1.8$--$300$~K in $0.5$~T magnetic field. No sign of magnetic ordering was observed down to $1.8$~K. The inverse susceptibility (Fig.~\ref{fig:M}) was fitted with the Curie-Weiss relation,
\begin{equation} 
\chi^{-1}= \frac{\rm C}{T-\theta_{\rm W}} ,
  \label{eqn:CW}
\end{equation}
where $\theta_{\rm W}$ is the Curie-Weiss temperature  and ${\rm C}$ is the Curie constant. This fit yielded: \mbox{$\theta_{\rm W} = 2.9$~K} and $\mu_{eff}$ = 7.54~$\mu_{\rm B}$. The effective moment closely matches that of Tm$^{3+}$ ion ($g_J\sqrt{J(J+1)}= 7.57$~$\mu_{\rm B}$). The magnetic isotherms of the compound were also measured at several temperatures. That for $2$~K displays a tendency to saturation above $3$~T (Fig.~\ref{fig:M}, inset), although they did not reach saturation value expected for Tm$^{3+}$ ion ($7 \mu_{\rm B}$) even after application of $14$~T magnetic field (\textit{cf}.~Fig.~\ref{fig:MH14}). 
The lattice parameter ($6.473$~\AA) and the susceptibility of our TmPdSb single crystals match well those reported earlier for poly-crystalline samples~\cite{malik1991magnetic}.

\end{document}